\documentclass[11pt]{article}
\usepackage{amsmath,amssymb,amscd}
\usepackage{amsthm,bbm}
\usepackage{mathtools}
\usepackage{float}
\usepackage[caption = false]{subfig}
\usepackage[final]{graphicx}
\usepackage[mathscr]{eucal}
\usepackage{color}
\usepackage{booktabs}
\usepackage{lineno}

\newtheorem{lemma}{Lemma}
\large\normalsize

\title{Average abundancy of cooperation in multi-player games with random payoffs}
\author{Dhaker Kroumi$^1$\footnote{Author for
correspondence, and e-mail: dhaker.kroumi@kfupm.edu.sa} and Sabin Lessard$^2$
\\$^1$Department of Mathematics and Statistics\\King Fahd University of Petroleum and Minerals\\Dhahran 31261, Saudi Arabia\\
$^2$Department of Mathematics and Statistics\\University of Montreal \\
 Montreal H3C 3J7, Canada\\
 }
\date{}
\begin{document}
\maketitle


\section*{Abstract}

We consider interactions between players in groups of size $d\geq2$ with payoffs that not only depend on the strategies used in the group but also fluctuate at random over time. An individual can adopt either cooperation or defection as strategy and the population is updated from one-time step to the next by a birth-death event according to a Moran model. Assuming recurrent symmetric mutation and payoffs with expected values, variances, and covariances of the same small order, we derive a first-order approximation of the average abundance of cooperation in the selection-mutation equilibrium. We show that increasing the variance of any payoff for defection or decreasing the variance of any payoff for cooperation increases the average abundance of cooperation. As for the effect of the covariance between any payoff for cooperation and any payoff for defection, we show that it depends on the number of cooperators in the group associated with these payoffs. We study in particular the public goods game, the stag hunt game, and the snowdrift game, all social dilemmas based on random benefit $b$ and cost $c$ for cooperation. We show that a decrease in the scaled variance of $b$ or $c$, or an increase in their scaled covariance, makes it easier for weak selection to favor the abundance of cooperation in the stag hunt game and the snowdrift game. The same conclusion holds for the public goods game except that the covariance of $b$ has no effect on the average abundance of $C$. On the other hand, increasing the scaled mutation rate or the group size can enhance or lessen the condition for weak selection to favor the abundance of $C$.


\noindent \textbf{Keywords and phrases}:   Moran model; Mutation-selection equilibrium; Evolution of cooperation; Dirichlet distribution; Public goods game; Stag hunt game; Snowdrift game

\noindent \textbf{Mathematics Subject Classification (2010)}: Primary 92D25; Secondary 60J70


\section{Introduction}

Evolutionary game theory assumes that the reproductive success of a strategy is not constant but depends on the frequencies of the different strategies (Maynard Smith and Price \cite{MP1973}, Maynard Smith \cite{M1982}, Hofbauer and Sigmund \cite{HS1988}). This reproductive success is a function of a mean payoff that depends on interactions between individuals, which in turn depend on the population state.

In continuous time, the mean payoff of a strategy has been defined as its growth rate. Then, in an infinitely large well-mixed population, the deterministic dynamics is described by the replicator equation (Taylor and Jonker \cite{TJ1978}, Zeeman \cite{Z1980}). Evolutionary concepts such as evolutionarily stability (Maynard Smith and Price \cite{MP1973}), continuous stability (Eshel \cite{E1983}) or convergence stability (Christiansen \cite{C1991}) were first studied in this framework (see, e.g., Taylor \cite{T1989}, Hofbauer and Sigmund \cite{HS1988}).
 
For evolutionary game dynamics in a finite population,  we have to resort to stochastic processes and probability concepts. Consider, for instance, a well-mixed population of fixed size $N\geq 2$ in discrete time, where each individual can use as strategy either $C$ for cooperation or $D$ for defection in a Prisoner's dilemma. With any update rule from one time step to the next and in the absence of mutation, the population state over time is described by a Markov chain that has two absorbing states corresponding to an all cooperating population and an all defecting population. Let $\rho_C$ (respectively, $\rho_D$) be the probability that a single individual of type $C$ (respectively, type $D$) among $N-1$ individuals of type $D$ (respectively, type $C$) generates a lineage forward in time that will take over the whole population.  Then, selection is said to favor the evolution of $C$ more than the evolution of $D$ if $\rho_C>\rho_D$ (Nowak \textit{et al.} \cite{NSTF2004}, Imhof and Nowak \cite{IN2006}). In the presence of symmetric mutation, the Markov chain is irreducible and, as a result, it possesses a stationary state called the mutation-selection equilibrium. If the average frequency of $C$ in this equilibrium is greater that the average frequency of $D$, then selection is said to favor the abundance of $C$ (Antal \textit{et al.} \cite{ANT2009}, see also Kroumi and Lessard \cite{KL2015a,KL2015b}).

The above game dynamics were considered first under the assumption of random pairwise interactions. Multi-player games were considered later on to take into account interactions within groups of any fixed size $d\geq2$. Kurokawa and Ihara \cite{KI2009}, for instance, studied the probability of ultimate fixation of a strategy given its initial frequency in this framework in the absence of the mutation, and Gokhale and Traulsen \cite{GT2011} its average abundance in the presence of recurrent mutation. In particular, an exact condition for weak selection to favor the abundance of a given strategy in a large population were deduced in the case $d=3$.

In all the above mentioned studies, the payoffs were supposed deterministic. This assumes a constant 
environment which is not generally the case in biological populations. Unknown risk of predation and variability in available resources, competition capabilities as well as birth and death rates (May \cite{M1973}, Kaplan \textit{et al.} \cite{KHH1990}, Lande \textit{et al.} \cite{LES2003}) are among multiple factors that motivate taking into account random fluctuations in evolutionary models.
Indeed, such fluctuations have an effect on evolutionary outcomes. Early studies on the effect of varying the selection coefficients between generations and varying the offspring numbers within generations in haploid as well as diploid population genetic models in the absence of mutation include Gillespie \cite{G1973,G1974}, Karlin and Levikson \cite{KL1974a}, Karlin and Liberman \cite{KL1974b} and Frank and Slatkin \cite{FS1990}). Extensions can be found in Starrfelt and Kokko \cite{SK2012}, Schreiber \cite{S2015} or Rychtar and Taylor \cite{RT2020}. Moreover, the fixation probability for a given type in a population whose size fluctuates dynamically was addressed in Lambert \cite{L2006}, Parsons and Quince \cite{PQa2007,PQb2007} and Otto and Whitlock \cite{OW1997}, among others, while Uecker and Hermisson \cite{UH2011} studied a population with temporal variation not only in its size but also in selection pressure. Competing populations distributed over habitat patches where environmental conditions fluctuate in time and space were considered too (Evans \textit{et al.} \cite{EHS2015}, Schreiber \cite{S2012}).

In evolutionary games, the payoffs may change at random over time. A stochastic version of the continuous-time replicator equation with a random noise added to the growth rate of every strategy was considered in Fudenberg and Harris \cite{FH1992}. More recently, the effect of stochastic changes in payoffs in two-player linear games in discrete time were studied with particular attention to stochastic local stability of fixation states and polymorphic equilibria in an infinite population (Zheng \emph{et al.} \cite{ZLLT2017, ZLLT2018}) and fixation probabilities of strategies in a finite population (Li and Lessard \cite{LL2020}, Kroumi \emph{et al.} \cite{KMLL2021}).

In a previous paper (Kroumi and Lessard \cite{KL2020}), we have studied the effects of randomness in payoffs on the average abundance of strategies under recurrent mutation in two-player games. Assuming a finite population in discrete time updated according to a Moran model. The payoffs for cooperation and defection in Prisoner's dilemmas, repeated or not, fluctuate over time such that their means, variances and covariances are of the same small order while higher moments are insignificant. In the mutation-selection equilibrium, we have shown that an increase in the variance of any payoff received by defection against cooperation or defection, or in their covariance, or a decrease in the variance of any payoff received by cooperation against cooperation or defection, or in their covariance, increases the average abundance of cooperation. Then, it is easier for weak selection to favor the abundance of cooperation. Moreover, increasing the scaled mutation rate can lessen or enhance the effect.

In this paper, we extend our analysis of average abundance with random payoffs 
to many-player games. Interactions occur within groups of any fixed size $d\geq2$ and the payoffs received by cooperators and defectors  are random variables that depend on the group composition. We study the average abundance of $C$ in the stationary state of the mutation-selection equilibrium to derive conditions for weak selection to favor the abundance of $C$. These conditions are examined in detail to understand the effect of the scaled mutation rate and the group size in different games and scenarios.

The remainder of this paper is organized as follows. In Section $2$, we formulate the model. In Section 3, we derive the average abundance of $C$ in the stationary state under symmetric mutation and weak selection. 
 In Section $4$, we examine in detail the particular case of payoffs with additive mean scaled cost and benefit for cooperation. In the next sections, we focus on three special social dilemmas, the public goods game in Section $5$, the stag hunt game in Section $6$, and the snowdrift game in Section $7$. We summarize the conclusions and interpretations of our results in Section $8$.

\section{Model}
Consider a population of a fixed size $N$. Each individual can be of only one of two types depending on the strategy used: $C$ for cooperation or $D$ for defection. Interactions between individuals occur within random groups of $d$ players. 
A cooperator (respectively, defector) receives a payoff $a_k$ (respectively, $b_k$) when it interacts with $k$ cooperators and $ d-k-1$ defectors.
The payoffs at a given time step are described in the following payoff array:
\begin{linenomath*} 
\begin{table}[!ht]  
  \centering
  \begin{tabular}{ccccc}
    \toprule
           &$0$ & $1$ & $\cdots$ & $d-1$\\ \midrule
  $C$ & $a_0$   & $a_1$ &  $\cdots$ & $a_{d-1}$  \\
 $D$ & $b_0$   & $b_1$ &  $\cdots$ & $b_{d-1}$  \\ \bottomrule
  \end{tabular}
\end{table}
\end{linenomath*} 
Here, we suppose that the payoffs are random variables whose first and second moments are given by
\begin{linenomath*} 
\begin{subequations}\label{momentspayoffs}
\begin{align}
&E\left[a_i\right]=\mu_{C,i}\delta+o(\delta),\\
&E\left[a_ia_j\right]=\sigma_{CC,ij}\delta+o(\delta),\\
&E\left[b_i\right]=\mu_{D,i}\delta+o(\delta),\\
&E\left[b_ib_j\right]=\sigma_{DD,ij}\delta+o(\delta),\\
&E\left[a_ib_j\right]=\sigma_{CD,ij}\delta+o(\delta),
\end{align}
\end{subequations}
\end{linenomath*} 
for $i,j=0,1,\ldots,d-1$. The parameter $\delta\geq0$, which corresponds to an intensity of selection, measures the order of the first and second moments of the payoffs. The parameters $\mu_{S_1,i}$ and $\sigma_{S_1S_2,ij}$ for $i,j=0,1,\ldots,d-1$ and $S_1, S_2= C$ or $D$ correspond to scaled means and covariances, respectively. In addition, all higher-order moments of the payoffs are assumed to be negligible compared to $\delta$, that is,
\begin{equation}
E\left[\prod_{i=0}^{d-1}a_i^{k_i}b_i^{l_i}\right]=o(\delta)
\end{equation}
as soon as $\sum_{i=0}^{d-1}(k_i+l_i)\geq 3$ for integers $k_i,l_i\geq0$ for $i=1, \ldots, d-1$. Moreover, the payoffs at any given time step are assumed to be independent of the payoffs at all other time steps.

Following interactions within groups, each individual accumulates a payoff $P$ that is translated into a reproductive fitness that takes the form
$f=1+P$.
Let $f_C(x)$ and $f_D(x)$ be the reproductive fitnesses of a cooperator and a defector, respectively, at a given time step when the frequency of $C$ in the population is $x=i/N$. Since groups of size $d$ are formed at random, 
we have
\begin{subequations}\label{payoffsCD}
\begin{align}
f_C(x)&=1+\sum_{k=0}^{d-1}\frac{\binom{i-1}{k}\binom{N-i}{d-k-1}}{\binom{N-1}{d-k-1}}a_k=1+P_C(x),\\
f_D(x)&=1+\sum_{k=0}^{d-1}\frac{\binom{i}{k}\binom{N-i-1}{d-k-1}}{\binom{N-1}{d-k-1}}b_k=1+P_D(x),
\end{align}
\end{subequations}
where $P_C(x)$ and $P_D(x)$ are the average payoffs to $C$ and $D$, respectively. 

The update of the population from one time step to the next is done through a single birth-death event according to a Moran model allowing for mutation (Moran \cite{M1958}, Ewens \cite{E2004}).
With probability proportional to its reproductive fitness, an individual is chosen to produce an offspring. With probability $1-u<1$, the offspring is an exact copy of its parent and, therefore, uses the same strategy. With the complementary probability $u>0$, the offspring is a mutant, in which case its strategy is chosen at random. More precisely, a mutant offspring adopts strategy $C$ with probability $1/2$ or strategy $D$ with probability $1/2$. In all cases, the offspring produced replaces an individual that is chosen at random in the population to die, possibly the parent of the offspring but not the offspring itself. This leads to the population state at the next time step.

The state space for the frequency of $C$ in the population at a given time step, represented by $X$, is $S=\{0,1/N,\ldots,(N-1)/N, 1\}$. The frequency of $C$ over all time steps is an aperiodic irreducible Markov chain on a finite state space. Owing to the ergodic theorem (see, e.g., Karlin and Taylor \cite{KT1975}), the chain tends to an equilibrium state, called the selection-mutation equilibrium, given by a unique stationary probability distribution, represented by  $\{\Pi^{\delta}(x)\}_{x\in S}$ where 
$\Pi^{\delta}(x)=\mathbb{P}^{\delta}(X=x)>0$ and $\sum\limits_{x\in S}\Pi^{\delta}(x)=1$.

Let $\mathbb{E}^{\delta}$ denote the expectation with respect to the stationary probability distribution if the intensity of selection is 
$\delta \geq 0$. The \emph{average abundance} of $C$ is defined as
\begin{equation}
\mathbb{E}^{\delta}[X]=\sum_{x\in \mathbf{S}}x\Pi^{\delta}(x).
\end{equation}
We say that \emph{weak selection favors the abundance of $C$} if its average abundance under weak enough selection exceeds what it would be under neutrality, that is,
\begin{equation} \label{favorabundance}
\mathbb{E}^{\delta}[X]>\mathbb{E}^{0}[X]=\frac{1}{2}
\end{equation}
for $\delta > 0$ small enough.
Here, $\mathbb{E}^{0}[X]$ represents the average abundance of $C$ under neutrality when $\delta=0$.


\section{Average abundance under symmetric mutation and weak selection}
From one time step to the next, the frequency of $C$ can increase by $1/N$, decrease by $1/N$, or remain the same. Let us denote this change by $\Delta X=X'-X$. We have $\Delta X=-1/N$ when a $D$ offspring replaces a $C$ individual. Let $T^-(x)$ be the conditional probability that $\Delta X=-1/N$ given that $X=x$. Then, we have
\begin{align}
T^-(x)&=\mathbb{P}^\delta\left[\Delta X=-\frac{1}{N}\,\Big{|} \, X=x\right]\nonumber\\
&=\Bigg[(1-u)E\left[\frac{(1-x)f_D(x)}{xf_C(x)+(1-x)f_D(x)} \right]+\frac{u}{2} \Bigg] x,
\end{align}
where  $E$ denotes an expectation with respect to the probability distribution of the payoffs.
Similarly, $\Delta X=1/N$ when a $C$ offspring replaces a $D$ individual, and the conditional probability of this event given that $X=x$ is
\begin{align}
T^+(x)&=\mathbb{P}^\delta\left[\Delta X=\frac{1}{N}\,\Big{|} \, X=x\right]\nonumber\\
&=\Bigg[(1-u)E\left[\frac{xf_C(x)}{xf_C(x)+(1-x)f_D(x)} \right]+\frac{u}{2} \Bigg] (1-x).
\end{align}
 Accordingly, the conditional expected change in the frequency of $C$ is
\begin{align}\label{S2-eq4}
\mathbb{E}_x^{\delta}\left[\Delta X\right]&=\mathbb{E}^{\delta}\left[\Delta X\, |\, X=x\right]\nonumber\\
&=\frac{1}{N}\left(T^+(x)-T^-(x)\right)\nonumber\\
&=\frac{u(1-2x)}{2N}+\frac{\delta}{N}(1-u)x(1-x)m(x)+o(\delta),
\end{align}
where
\begin{equation}
m(x)=E\left[\frac{f_C(x)-f_D(x)}{xf_C(x)+(1-x)f_D(x)}\right].
\end{equation}

Multiplying both sides in (\ref{S2-eq4}) by $\Pi^{\delta}(x)$ and summing up over all states $x$ in $S$ lead to
\begin{equation}\label{S2-eq6}
\mathbb{E}^{\delta}\left[\Delta X\right]=\frac{u}{2N}\left[1-2\mathbb{E}^{\delta}[X]\right]+\frac{\delta}{N}(1-u)\mathbb{E}^{\delta}\left[X(1-X)m(X)\right]+o(\delta).
\end{equation}
In the stationary state, the frequency of $C$ in the population keeps a constant expected value, that is,
\begin{equation}\label{s2-eq7}
\mathbb{E}^{\delta}\left[\Delta X\right]=0.
\end{equation}
Therefore, (\ref{S2-eq6}) yields
\begin{equation}\label{s2-eq8}
\mathbb{E}^{\delta}[X]=\frac{1}{2}+\frac{\delta(1-u)}{u}\mathbb{E}^{\delta}\left[X(1-X)m(X)\right]+o(\delta).
\end{equation}
Then, using 
\begin{equation}\label{s2-eq9}
\mathbb{E}^{\delta}\left[X(1-X)m(X)\right]=\mathbb{E}^{0}\left[X(1-X)m(X)\right]+O(\delta),
\end{equation}
we obtain the first-order approximation
\begin{equation}\label{s2-eq10}
\mathbb{E}^{\delta}[X]\approx\frac{1}{2}+\frac{\delta(1-u)}{u}\mathbb{E}^{0}\left[X(1-X)m(X)\right]
\end{equation}
for the expected frequency of $C$ in the selection-mutation equilibrium. Returning to (\ref{favorabundance}), we have that weak selection favors the abundance of $C$ if
\begin{equation}\label{main}
\mathbb{E}^{0}\left[X(1-X)m(X)\right]>0.
\end{equation}

In the remainder of this paper, we consider a large population size, in which case we have
\begin{align}\label{S2-eq5}
m(x)&=\sum_{k=0}^{d-1}\binom{d-1}{k}x^k(1-x)^{d-k-1}\left(\mu_{C,k}-\mu_{D,k}\right)\nonumber\\
&\quad+\sum_{k,l=0}^{d-1}\binom{d-1}{k}\binom{d-1}{l}x^{k+l}(1-x)^{2d-k-l-2}\nonumber\\
&\quad\quad \times \Big[x(\sigma_{CD,kl}-\sigma_{CC,kl})+(1-x)(\sigma_{DD,kl}-\sigma_{CD,kl})\Big].
\end{align}
See Appendix $A$.


Now, let us define
\begin{equation}\label{psi}
\psi_n^k=\mathbb{E}^{0}\left[X^k(1-X)^{n-k}\right]
\end{equation}
for $k=0, 1, \ldots, n$. This is the probability that among $n$ individuals drawn at random with replacement in a neutral population at equilibrium, $k$ of them in particular are of type $C$, while the other $n-k$ individuals are of type $D$. Note that
$\psi^k_n=\psi^{n-k}_n$
since mutation is symmetric.

In a neutral population in the limit of a large population size $N$ with $N^2/2$ birth-death events as unit of time, each pair of lineages coalesces backward in time at rate $1$ independently of all others according to Kingman's coalescent 
 (Kingman\cite{K1982}). Besides, each lineage mutates at the scaled rate $\theta= \lim_{N\rightarrow\infty} Nu/2$ independently of all others and independently of the coalescent process (see, e.g., Ewens\cite{E2004}, p. 340). This implies that the expected value in (\ref{psi}) in the limit of a large population size corresponds to a moment of a Dirichlet distribution (Ewens\cite{E2004}, p. 195). This leads to the following key lemma (see Appendix B for a proof).

\begin{lemma} In a large population, we have the approximation
\begin{align}
\psi_n^k&\approx \frac{\prod_{i=1}^{k}(\theta+i-1)\prod_{j=1}^{n-k}(\theta+j-1)}{\prod_{l=1}^{n}(2\theta+l-1)}
\end{align}
for $k=0, 1, \ldots, n$, where $\theta= \lim_{N\rightarrow\infty} Nu/2$ is a scaled mutation rate.
\end{lemma}

Note that we have the approximation 
\begin{align}\label{approximationsmall}
\psi_n^k=\left\{
    \begin{array}{ll}
       \frac{\theta(k-1)! (n-k-1)!}{2(n-1)!}& \mbox{if } k=1,2,\ldots,n-1, \\
       \frac{1}{2}-\frac{\theta}{2}H_{n-1} & \mbox{if } k=0,n,
    \end{array}
\right.
\end{align}
when $\theta$ is small and terms of order $o(\theta)$ are neglected,
and the approximation
\begin{equation}\label{approximationlarge}
\psi_n^k\approx \frac{1}{2^{n}}
\end{equation}
for $k=0,1,\ldots,n$ when $\theta$ is large and terms of order $O(\theta^{-1})$ are neglected. Here, $H_k=\sum_{i=1}^{k}1/i$ denotes the $k$-th harmonic number (Conway and Guy \cite{CG1995}).

Using (\ref{S2-eq5}) and (\ref{psi}), the expected value on the left-hand side of (\ref{main}) can be expressed as
\begin{align}\label{mainresult}
\mathbb{E}^{0}\left[X(1-X)m(X)\right]&=\sum_{k=0}^{d-1}\binom{d-1}{k}\psi_{d+1}^{k+1}\left(\mu_{C,k}-\mu_{D,k}\right)+\sum_{k,l=0}^{d-1}\binom{d-1}{k}\binom{d-1}{l}\nonumber\\
&\quad \times \left[-\psi_{2d+1}^{k+l+2}\left(\sigma_{CC,kl}-\sigma_{CD,kl}\right)+\psi_{2d+1}^{k+l+1}\left(\sigma_{DD,kl}-\sigma_{CD,kl}\right)\right].
\end{align}
In this expression, the coefficient of $\sigma_{CC,kl}$ is always negative, while 
the coefficient of $\sigma_{DD,kl}$ is always positive, for $k,l=0, 1,\ldots,d-1$.  As for the effect of a change in any covariance between a payoff to $C$ and a payoff to $D$, $\sigma_{CD,kl}$, on average abundance, it depends on the sign of $\psi_{2d+1}^{k+l+2}-\psi_{2d+1}^{k+l+1}$, for $k,l =0, 1, \ldots, d-1$. Owing to the above lemma, we have the approximation
\begin{equation}
\psi_{2d+1}^{k+l+2}-\psi_{2d+1}^{k+l+1}\approx \frac{(k+l+1-d)\prod_{i=1}^{k+l+2}(\theta+i-1)\prod_{j=1}^{2d-k-l-1}(\theta+j-1)}{\prod_{l=1}^{2d+1}(2\theta+l-1)}
\end{equation} 
if the population size is large enough. This approximation is positive if $k+l>d-1$, negative if $k+l<d-1$, and null if $k+l=d-1$. 

Therefore, the following conclusion ensues.

\paragraph{Result 1}\textit{
Under weak selection, decreasing any covariance $\sigma_{CC,kl}$ between two payoffs to $C$, or increasing any covariance $\sigma_{DD,kl}$ between two payoffs to $D$, increases the average abundance of $C$. In other words, less uncertainty in the payoffs to $C$ or more uncertainty in the payoffs to $D$ makes it easier for weak selection to favor the abundance of $C$. Moreover, increasing any covariance $\sigma_{CD,kl}$ between a payoff to $C$ and a payoff to $D$ for $k+l>d-1$, or decreasing it for $k+l<d-1$, increases the average abundance of $C$, while increasing it or decreasing it for  $k+l=d-1$ has no effect on the average abundance of $C$.}\\

Note that, if the population size is large enough and all the payoffs are constant, so that all the covariances vanish, then condition (\ref{main}) for weak selection to favor the abundance of $C$ reduces to
\begin{equation}
\sum_{k=0}^{d-1}\binom{d-1}{k}\Gamma(\theta+k+1)\Gamma(\theta+d-k)\left(\mu_{C,k}-\mu_{D,k}\right) > 0,
\end{equation}
where $\Gamma(\beta + 1)=\beta \Gamma(\beta)$ for $\beta >0$. This generalizes a result of Gokhale and Traulsen \cite{GT2011} for $d=3$.


\section{Additive scaled mean cost and benefit}
In this section, we focus on a particular case where the scaled mean payoffs to $C$ and $D$ are given by 
\begin{subequations}
\begin{align}
\mu_{C,k}&=\frac{k}{d-1}\mu_{b}-\mu_{c},\\
\mu_{D,k}&=\frac{k}{d-1}\mu_{b},
\end{align}
\end{subequations}
respectively, for $k=0, 1, \ldots,d-1$. This corresponds to a public goods game where a cooperator pays a mean scaled cost $\mu_c$, while the scaled mean benefit of cooperation $\mu_b$ is distributed equally among the other $d-1$ members of the group.

Under constant payoffs, that is, $\sigma_{S_1S_2,kl}=0$ for all $S_1,S_2=C,D$ and $k=0,1,\ldots,d-1$, weak selection never favors the abundance of $C$ since then the espression (\ref{mainresult}) reduces to the first summation given by 
\begin{equation}\label{firstsummation}
\sum_{k=0}^{d-1}\binom{d-1}{k}\psi_{d+1}^{k+1}\left(\mu_{C,k}-\mu_{D,k}\right)=-\mu_c\sum_{k=0}^{d-1}\binom{d-1}{k}\psi_{d+1}^{k+1}= -\mu_c\psi_{2}^{1},
\end{equation}
which cannot be positive.
Here, we have used the identity
\begin{align}\label{identity1}
\sum_{k=0}^{n}\binom{n}{k}\psi_{n+i}^{k+j}&=\sum_{k=0}^{n}\binom{n}{k}\mathbb{E}^{0}\left[X^{k+j}(1-X)^{n+i-k-j}\right]\nonumber\\
&=\mathbb{E}^{0}\left[X^{j}\left(1-X\right)^{i-j}\sum_{k=0}^{n}\binom{n}{k}X^{k}\left(1-X\right)^{n-k}\right]\nonumber\\
&=\mathbb{E}^{0}\left[X^{j}\left(1-X\right)^{i-j}\right]=\psi^{j}_{i},
\end{align}
for $0\leq j\leq i$.

In the remainder of this section, we will show that introducing uncertainty in the payoffs to $D$ can make it possible for weak selection to favor the abundance of $C$.
\subsection{Case 1}
In this subsection, we suppose that all the covariances between any two payoffs to $C$ and all the covariances between any payoff to $C$ and any payoff to $D$ are insignificant, that is, $\sigma_{CC,kl}=0$ and $\sigma_{CD,kl}=0$ for $k,l=0,1,\ldots,d-1$. Moreover, we suppose that all the covariances between any two payoffs to $D$ are insignificant except for a certain integer $k_0$ between $0$ and $d-1$. More precisely, we have $\sigma_{DD,kl}=0$ for $(k,l)\not=(k_0,k_0)$ and 
$\sigma_{DD,k_0k_0}=\sigma^2 >0$.

 In this case, the second summation in (\ref{mainresult}) reduces to $\sigma^2\binom{d-1}{k_0}^2\psi_{2d+1}^{2k_0+1}$. Then, using the expression of the first summation given in (\ref{firstsummation}), condition (\ref{main}) for weak selection to favor the abundance of $C$ takes the form
\begin{equation}\label{case1}
\frac{\sigma^2}{\mu_c}>\left(\frac{\sigma^2}{\mu_c}\right)^*=\frac{\psi_{2}^{1}}{\binom{d-1}{k_0}^2\psi_{2d+1}^{2k_0+1}}.
\end{equation}
See Figure 2 for graphics of this threshold value in a large population with respect to the scaled mutation rate $\theta$ for $d=2,3,4,5$ and $k_0=0,1,\ldots,d-1$ using the approximation that is given in Lemma 1. 

Using the approximation in (\ref{approximationsmall}) when $\theta$ is small, we get in this case
\begin{equation}
\left(\frac{\sigma^2}{\mu_c}\right)^*\approx\frac{(2k_0+1)\binom{2d}{2k_0+1}}{\binom{d-1}{k_0}^2}.
\end{equation}
For any $d\geq2$, the value of this threshold is an increasing function of $k_0$ since we have
\begin{equation}
\frac{\frac{\left(2\left(k_0+1\right)+1\right)\binom{2d}{2(k_0+1)+1}}{\binom{d-1}{k_0+1}^2}}{\frac{(2k_0+1)\binom{2d}{2k_0+1}}{\binom{d-1}{k_0}^2}}=\frac{(k_0+1)(2d-2k_0-1)}{(2k_0+1)(d-k_0-1)}=\frac{2k_0d-2k_0^2-3k_0+2d-1}{2k_0d-2k_0^2-3k_0+d-1}>1.
\end{equation}
We conclude that the best scenario for the abundance of $C$ to be favored by weak selection is when $k_0=0$, in which case the threshold value is minimum at $2d$, while the worst scenario is when $k_0=d-1$, with the threshold value reaching its maximum $2d(2d-1)$.
Note that the threshold value tends to $\infty$ as $d\rightarrow\infty$. Therefore, if the group size $d$ is large enough and the scaled mutation rate $\theta$ low enough, weak selection cannot favor the abundance of $C$.

\begin{center}
\begin{figure}
\includegraphics[height=10cm, width=16cm]{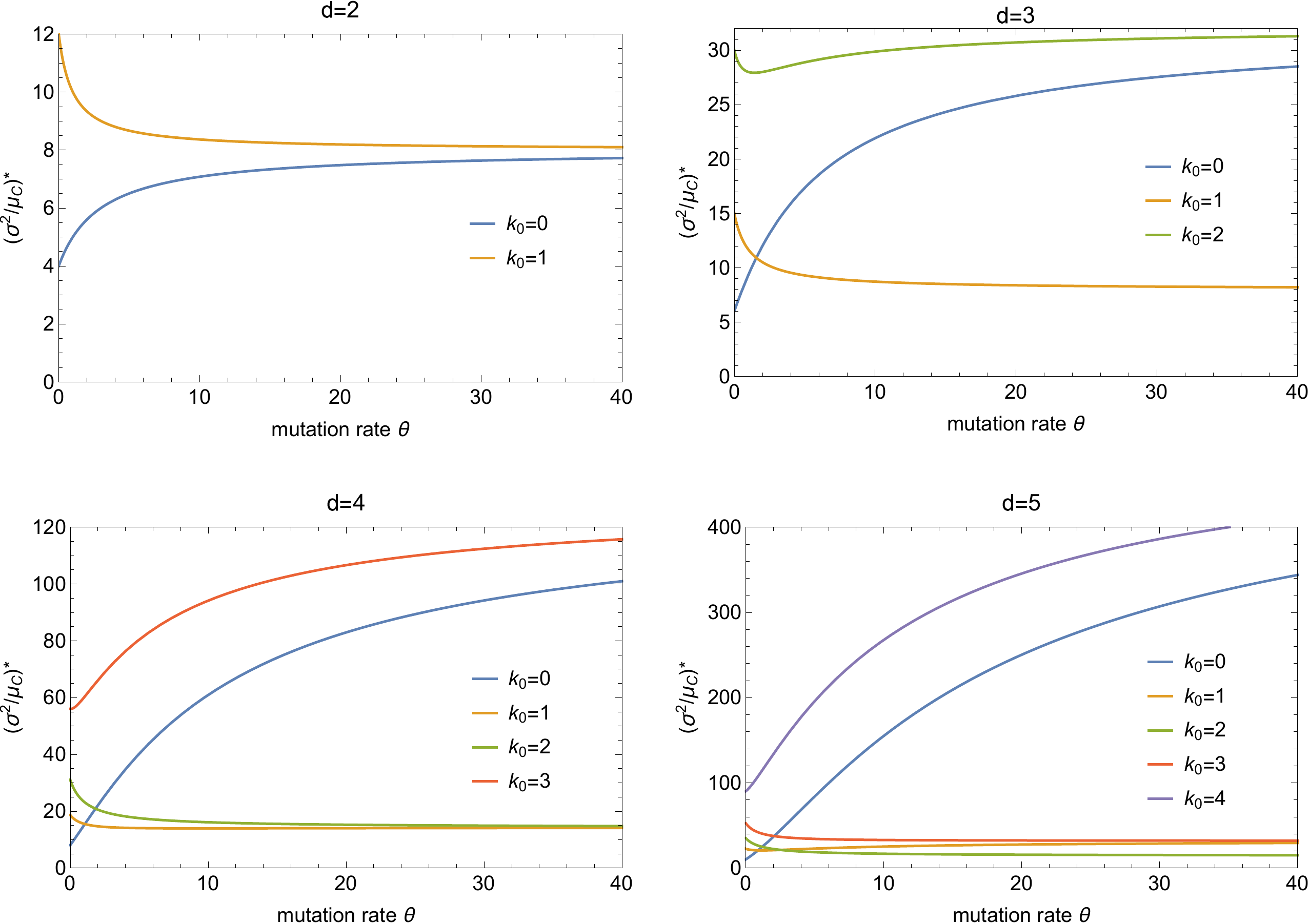}
\caption{Curves of the threshold value $(\sigma^2/\mu_c)^*$ that $\sigma^2/\mu_c$ must exceed for weak selection to favor the abundance of $C$ for $d=2,3,4,5$ and $k_0=0,1,\ldots,d-1$ with respect to the scaled mutation rate $\theta$ in a large population. Under a low scaled mutation rate, the best scenario for the  abundance of $C$ is when $k_0=0$, while the worst scenario is when $k_0=d-1$. Under a high scaled mutation rate, the worst scenario is when $k_0=0$ or $d-1$, and the best scenario when $k_0=(d-1)/2$ if $d$ is odd, or 
$k_0=d/2-1$ or $k_0=d/2$ if $d$ is even. Note that the scaled mutation rate can increase or decrease the threshold depending on $k_0$. Increasing the group size $d$ increases the threshold value $(\sigma^2/\mu_c)^*$, which makes it more difficult for weak selection to favor the abundance of $C$.}
\label{figure2}
\end{figure}
\end{center}

On the other hand, using the approximation in (\ref{approximationlarge}) in the case where $\theta$ is large, we have
\begin{equation}
\left(\frac{\sigma^2}{\mu_c}\right)^*\approx \frac{2^{2d-1}}{\binom{d-1}{k_0}^2}\,,
\end{equation} 
which reaches its maximum when $k_0=0$ or $d-1$, and its minimum when $k_0=(d-1)/2$ if $d$ is odd, or when $k_0=d/2-1$ or $k_0=d/2$ if $d$ is even. Note that, as $d\rightarrow\infty$, the threshold value tends to $\infty$, in which case weak selection cannot favor the abundance of $C$. 

\subsection{Case 2}
We suppose now that all the covariances between any two payoffs to $C$ and all the covariances between any payoff to $C$ and any payoff to $D$ are insignificant, that is, $\sigma_{CC,kl}=\sigma_{CD,kl}=0$ for $k,l=0,1,\ldots,d-1$. Moreover, we suppose that all the covariances between any two payoffs to $D$ are of the same positive order, that is, $\sigma_{DD,kl}=\sigma^2>0$ for $k,l=0,1,\ldots,d-1$. This is the case, for instance, when all payoffs to $D$ are perfectly positively correlated.

In this case, the second summation in (\ref{mainresult}) takes the form
\begin{equation}\label{sec4-case2-eq1}
\sigma^2\sum_{k,l=0}^{d-1}\binom{d-1}{k}\binom{d-1}{l}\psi_{2d+1}^{k+l+1}=\sigma^2\psi^{1}_{3}.
\end{equation}
Here, we have used the identity
\begin{align}\label{identity2}
&\sum_{k,l=0}^{d-1}\binom{d-1}{k}\binom{d-1}{l}\psi_{2d+i}^{k+l+j}\nonumber\\
&=\sum_{k,l=0}^{d-1}\binom{d-1}{k}\binom{d-1}{l}\mathbb{E}^{0}\left[X^{k+l+j}\left(1-X\right)^{2d+i-k-l-j} \right]\nonumber\\
&=\mathbb{E}^{0}\left[X^{j}\left(1-X\right)^{i+2-j}\Bigg(\sum_{k=0}^{d-1}\binom{d-1}{k}X^{k}\left(1-X\right)^{d-1-k}\Bigg)^2\right]\nonumber\\
&=\mathbb{E}^{0}\left[X^{j}\left(1-X\right)^{i+2-j}\right]=\psi_{i+2}^{j},
\end{align}
for $0\leq j\leq i+2$.
 Inserting the expressions (\ref{firstsummation}) and (\ref{sec4-case2-eq1})  in (\ref{s2-eq10}) and using Lemma $1$, the first-order approximation of the average abundance of $C$ becomes
\begin{equation}\label{sec4-case2-eq2}
\mathbb{E}^{\delta}[X]\approx\frac{1}{2}+\frac{\delta N}{8(2\theta+1)}\left(\sigma^2-2\mu_c\right).
\end{equation}
Accordingly, condition (\ref{main}) for weak selection to favor the abundance of $C$ can be written as
\begin{equation}\label{sec4-case2-eq3}
\frac{\sigma^2}{\mu_c}>2\, .
\end{equation}
This condition does not depend on $\theta$ nor on $d$. If it is satisfied, then weak selection favors the abundance of $C$ for any scaled mutation rate $\theta>0$ and any group size $d\geq2$. 

It is clear from (\ref{sec4-case2-eq2}) that increasing the scaled mutation rate will decrease the average abundance of $C$ if $\sigma^2>2\mu_c$, and increase it if $\sigma^2<2\mu_c$. On the other hand, the group size $d$ has no effect on the average abundance of $C$.

\section{Public goods game}
In this section and the next two ones, we are interested in classical social dilemmas where cooperation incurs a random cost $c>0$ but provides a random benefit $b>c$ in groups of size $d$. Moreover, we assume that
\begin{subequations}
\begin{align} 
 E[b]&=\mu_{b}\delta+o(\delta),\\
 E[c]&=\mu_{c}\delta+o(\delta),\\
 E[b^2]&=\sigma^2_b\delta+o(\delta),\\
E[c^2]&=\sigma^2_c\delta+o(\delta),\\
E[bc]&=\sigma_{bc}\delta+o(\delta)
\end{align}
\end{subequations}
and
\begin{equation}
E\left[b^ic^j\right]=o(\delta),
\end{equation}
for any non-negative integers $i$ and $j$ such that $i+j\geq3$.
 
We consider first a linear public goods game in which the benefit of cooperation $b$ by an individual at a cost $c$ is distributed equally among the $d-1$ other individuals in the same group and all effects of cooperation are additive (Hamburger \cite{H1973}, Fox and Guyer \cite{FG1978}, Nowak and Sigmund \cite{NS1990}, Wild and Traulsen \cite{WT2007}). In this case, the payoffs to $C$ and $D$ for an individual in interaction with $k$ cooperators and $d-k-1$ defectors are 
\begin{subequations}
\begin{align}
a_k&=\frac{k}{d-1}b-c,\\
b_k&=\frac{k}{d-1}b,
\end{align}
\end{subequations}
whose scaled means are given by
\begin{subequations}
\begin{align}
\mu_{C,k}&=\frac{k}{d-1}\mu_{b}-\mu_{c},\\
\mu_{D,k}&=\frac{k}{d-1}\mu_{b},
\end{align}
\end{subequations}
and scaled variances and covariances by 
\begin{subequations}
\begin{align}
\sigma_{CC,kl}&=\frac{kl}{(d-1)^2}\sigma^2_b+\sigma^2_c-\frac{k+l}{d-1}\sigma_{bc},\\
\sigma_{CD,kl}&=\frac{kl}{(d-1)^2}\sigma^2_b-\frac{k}{d-1}\sigma_{bc},\\
\sigma_{DD,kl}&=\frac{kl}{(d-1)^2}\sigma^2_b,
\end{align}
\end{subequations}
for $k,l=0, 1, \ldots,d-1$.
Then, the first summation in (\ref{mainresult}) is the same as in (\ref{firstsummation}).
On the other hand, using the identity $\frac{l}{d-1}\binom{d-1}{l}=\binom{d-2}{l-1}$, we obtain 
\begin{align}
&\sum_{k,l=0}^{d-1}\binom{d-1}{k}\binom{d-1}{l}\frac{l}{d-1}\psi_{2d+1}^{k+l+i}\nonumber\\
&=\sum_{k=0}^{d-1}\sum_{l=1}^{d-1}\binom{d-1}{k}\binom{d-2}{l-1}\mathbb{E}^{0}\left[X^{k+l+i}\left(1-X\right)^{2d+1-k-l-i}\right]\nonumber\\
&=\mathbb{E}^{0}\left[X^{i+1}(1-X)^{3-i}\sum_{k=0}^{d-1}\binom{d-1}{k}X^{k}\left(1-X\right)^{d-1-k}\sum_{l=0}^{d-2}\binom{d-2}{l}X^{l}\left(1-X\right)^{d-2-l}\right]\nonumber\\
&=\mathbb{E}^{0}\left[X^{i+1}(1-X)^{3-i}\right]=\psi_{4}^{i+1},
\end{align}
for $i=0,1,2,3$. Then, the second summation in (\ref{mainresult}) can be written as
\begin{equation}\label{publicgoods1}
\begin{split}
&\sum_{k,l=0}^{d-1}\binom{d-1}{k}\binom{d-1}{l}\left[\left(\frac{l}{d-1}\psi_{2d+1}^{k+l+2}+\frac{k}{d-1}\psi_{2d+1}^{k+l+1}\right)\sigma_{bc}-\psi_{2d+1}^{k+l+2}\sigma^2_{c}\right]\\
&=\left(\psi^2_4+\psi^3_4\right)\sigma_{bc}-\psi^2_3\sigma^2_c\\
&=\psi^2_3\left(\sigma_{bc}-\sigma^2_c\right).
\end{split}
\end{equation}
In the last passage, we have used the identity
\begin{equation}\label{identity3}
\psi^i_n+\psi^{i+1}_n=
\psi^i_{n-1}.
\end{equation}
Inserting (\ref{firstsummation}) and (\ref{publicgoods1}) in (\ref{s2-eq10}) and using Lemma $1$, the first-order approximation of the average abundance of $C$ is given by
\begin{equation}\label{publigoods}
\mathbb{E}^{\delta}[X]\approx\frac{1}{2}+\frac{\delta N}{8(2\theta+1)}\left(\sigma_{bc}-\sigma^2_c-2\mu_c\right).
\end{equation}
\paragraph{Result 2}\textit{
For the public goods game, decreasing the variance of the cooperation cost $c$, $\sigma^2_c$, or increasing the covariance between the cost $c$ and the benefit $b$, $\sigma_{bc}$, increases the average abundance of $C$. Neither the scaled mutation rate $\theta$ nor the group size $d$ has any effect on the condition for weak selection to favor the abundance of $C$ given by
\begin{equation}
\sigma_{bc}-\sigma^2_{c}>2\mu_c.
\end{equation}
}
Note that increasing the scaled mutation rate will decrease the average abundance of $C$ if $\sigma_{bc}-\sigma^2_{c}>2\mu_c$, and increase it if $\sigma_{bc}-\sigma^2_{c}<2\mu_c$. Therefore, increasing the scaled mutation rate will increase or decrease the average abundance of $C$ without changing the strategy whose abundance is favored by weak selection. If $b$ and $c$ are uncorrelated, that is, $\sigma_{bc}=0$, then weak selection cannot favor the abundance of $C$.

\section{Stag hunt game}
In a stag hunt game, an individual in a group of size $d$ receives a benefit $b$ only if all the individuals in the group cooperate, each one at a cost $c$ (Skyrms \cite{S2004}, Pacheco \textit{et al.} \cite{PSSS2009}). Then, the payoffs to $C$ are given by $a_k=-c$ if $k=0,1,\cdots,d-2$ and $a_{d-1}=b-c$, while the payoffs to $D$ are $b_l=0$ for $l=0,1,\ldots,d-1$. In this case, the scaled means of the payoffs to $C$ or $D$ according to the numbers of cooperating partners in the same group are given by
\begin{equation}
\mu_{C,k}= \left\{
    \begin{array}{ll}
       -\mu_{c}& \mbox{if } k < d-1, \\
       \mu_{b}-\mu_{c} & \mbox{if } k=d-1,
    \end{array}
\right.
\end{equation}
and 
$\mu_{D,l}=0$, and
the scaled variances and covariances by
\begin{equation}
\sigma_{CC,kl}= \left\{
    \begin{array}{ll}
       \sigma^2_{c}& \mbox{if } k < d-1 \mbox{ and } l < d-1,\\
      \sigma^2_{c}-\sigma_{bc}& \mbox{if } k=d-1 \mbox{ and } l<d-1\mbox{ or }k<d-1 \mbox{ and } l=d-1,\\
      \sigma^2_{c}-2\sigma_{bc}+\sigma^2_{b}& \mbox{if } k=l=d-1,
    \end{array}
\right.
\end{equation}
and $\sigma_{CD,kl}=\sigma_{DD,kl}=0$,
for $k,l=0, 1, \ldots, d-1$.

In this case, by using the identity (\ref{identity1}) and the fact that $\psi_{n}^{k}=\psi_{n}^{n-k}$, the first summation in (\ref{mainresult}) becomes
\begin{equation}\label{stag-eq1}
\psi_{d+1}^{d}\mu_{b}-\sum_{k=0}^{d-1}\binom{d-1}{k}\psi_{d+1}^{k+1}\mu_{c}=\psi_{d+1}^{1}\mu_{b}-\psi_{2}^{1}\mu_c.
\end{equation}
Similarly, using the  identities (\ref{identity1}) and (\ref{identity2}), the second summation in (\ref{mainresult}) takes the form
\begin{align}\label{stag-eq2}
&-\psi_{2d+1}^{2d}\sigma^2_{b}-\sum_{k,l=0}^{d-1}\binom{d-1}{k}\binom{d-1}{l}\psi_{2d+1}^{k+l+2}\sigma^2_{c}+2\sum_{k=0}^{d-1}\binom{d-1}{k}\psi_{2d+1}^{d+k+1}\sigma_{bc}\nonumber\\
&=-\psi_{2d+1}^{2d}\sigma^2_{b}-\psi_{3}^{2}\sigma^2_{c}+2\psi_{d+2}^{d+1}\sigma_{bc}=-\psi_{2d+1}^{1}\sigma^2_{b}-\psi_{3}^{1}\sigma^2_{c}+2\psi_{d+2}^{1}\sigma_{bc}.
\end{align}
Inserting (\ref{stag-eq1}) and (\ref{stag-eq2}) in (\ref{s2-eq10}), the average abundance of $C$ up to the first-order with respect to $\delta$ can be approximated as
\begin{equation}\label{stag-eq3}
\mathbb{E}^{\delta}[X]\approx\frac{1}{2}+\frac{\delta N}{2\theta}\left(\psi_{d+1}^{1}\mu_{b}-\psi_{2}^{1}\mu_c-\psi_{2d+1}^{1}\sigma^2_{b}-\psi_{3}^{1}\sigma^2_{c}+2\psi_{d+2}^{1}\sigma_{bc}\right).
\end{equation}
As a result, condition (\ref{main}) for weak selection to favor the abundance of $C$ can be reduced to
\begin{equation}\label{stag-eq6}
\psi_{d+1}^{1}\mu_{b}-\psi_{2}^{1}\mu_c-\psi_{2d+1}^{1}\sigma^2_{b}-\psi_{3}^{1}\sigma^2_{c}+2\psi_{d+2}^{1}\sigma_{bc}>0.
\end{equation}
This allows us to state our next result.

\paragraph{Result 3}\textit{
For the stag hunt game, decreasing the variance of the cost $c$, $\sigma^{2}_c$, or the variance of the benefit $b$, $\sigma^{2}_b$, or increasing their covariance, $\sigma_{bc}$, increases the average abundance of $C$ for any scaled mutation rate $\theta>0$ and any group size $d\geq2$.}\\

The next point of interest is the effect of the  group size $d$ on the condition for weak selection to favor the abundance of $C$.
Note that $\psi_{n}^{1}$ is decreasing as a function of $n$. Then, increasing the group size $d$ decreases the weights of $\mu_b$, $\sigma_{bc}$, and $\sigma^2_{b}$ on the average abundance of $C$, while the weights of $\mu_c$ and $\sigma^2_c$ remain the same. Using  
$\ln(1-x)\leq-x$
for $0\leq x <1$, we obtain
\begin{equation}\label{stag-eq8}
\begin{split}
\prod_{j=1}^{n}\left(\frac{\theta+j-1}{2\theta +j-1}\right)=\exp\left\{\sum_{j=1}^{n}\ln\left(1-\frac{\theta}{2\theta+j-1}\right)\right\}\leq
\exp\left\{-\sum_{j=1}^{n}\frac{\theta}{2\theta+j-1}\right\},
\end{split}
\end{equation}
from which
\begin{equation}\label{stag-eq9}
\lim_{n\rightarrow\infty}\prod_{i=1}^{n}\left(\frac{\theta+i}{2\theta+i}\right)=0,
\end{equation}
for any scaled mutation rate $\theta>0$.
On the other hand, using lemma $1$, we have
\begin{align}\label{stag-eq9bis}
0\leq \psi_n^k\approx \frac{\prod_{i=1}^{k}(\theta+i-1)\prod_{j=1}^{n-k}(\theta+j-1)}{\prod_{l=1}^{n}(2\theta+l-1)}
\leq  \frac{(\theta+k-1)^k}{(2\theta+n-k)^k}\prod_{j=1}^{n-k}\left(\frac{\theta+j-1}{2\theta+j-1}\right)
\end{align}
 for $k=0, 1, \ldots, n$. In particular, for a group size $d$ large enough, we have 
$\psi_{d+1}^{1}, \psi_{2d+1}^{1}, \psi_{d+2}^{1} \approx 0$,
and the approximation for the average abundance of $C$ given by (\ref{stag-eq3}) becomes
\begin{equation}
\mathbb{E}^{\delta}[X]\approx\frac{1}{2}-\frac{\delta N}{8(2\theta+1)}\left(2\mu_c+\sigma^2_c\right).
\end{equation}
This is always less than $1/2$ for any scaled mutation rate $\theta>0$. This means that weak selection cannot favor the abundance of $C$ in this case. Note that increasing the scaled mutation rate will increase the abundance of $C$ without changing the strategy whose abundance is favored, which can only be strategy $D$.

\paragraph{Result 4}\textit{
Increasing the group size $d$ in a large population makes it more difficult for weak selection to favor the abundance of $C$. If the group size is large enough, weak selection can only favor the abundance of $D$. This is true for any scaled mutation rate $\theta>0$.}\\

As for the effect of the scaled mutation on condition (\ref{stag-eq6}) in a large population, it is easy to see from lemma 1 that the coefficients of $\mu_b$, $\sigma^2_b$ and $\sigma_{bc}$ are decreasing with respect to $\theta$ for any $d\geq 2$.
 In the limit of a low scaled mutation rate, this condition takes the form
\begin{equation}\label{stag-eq10}
\frac{1}{d}\mu_b-\mu_c-\frac{1}{2d}\sigma^2_b-\frac{1}{2}\sigma^2_c+\frac{2}{d+1}\sigma_{bc}>0,
\end{equation}
while it reduces to
\begin{equation}\label{stag-eq11}
\frac{1}{2^{d-1}}\mu_b-2\mu_c-\frac{1}{2^{2d-1}}\sigma^2_b-\sigma^2_c+\frac{1}{2^d}\sigma_{bc}>0
\end{equation}
in the limit of a high scaled mutation rate. Note that with interactions in large groups, that is, as $d\rightarrow\infty$, conditions (\ref{stag-eq10}) and (\ref{stag-eq11}) are never satisfied, which means that weak selection cannot favor the abundance of $C$.

\section{Snowdrift game}
In a snowdrift game, the cost of a collective effort $c>0$ is distributed equally among the cooperators in the same group, while everyone in the group receives a benefit $b$ if there is at least one cooperator in the group (Zheng \textit{et al.} \cite{ZYCH2007}, Souza \textit{et al.} \cite{SPS2009}, Santos and Pacheco \cite{SP2011}). 
In this case, the payoffs to $C$ are $a_k=b-c/(k+1)$ for $k=0,1,\ldots,d-1$, while the payoffs to $D$ are
$b_l=b$ for $l=1,2,\ldots,d-1$, and $b_0=0$. Note that the cost for each cooperator is a non-linear decreasing function with respect to the number of cooperators in the group.

In this case, the scaled means, variances and covariances of the payoffs to an individual of type $C$ and $D$ according to the numbers of cooperating partners in the same group are given by
\begin{subequations}\label{snowdrift-eq0}
\begin{align}
\mu_{C,k}&=\mu_{b}-\frac{\mu_{c}}{k+1},\\
\mu_{D,k}&= \mu_{b}\mathbf{1}_{\{k\not=0\}},\\
\sigma_{CC,kl}&=\sigma^2_b-\left(\frac{1}{k+1}+\frac{1}{l+1}\right)\sigma_{bc}+\frac{\sigma^2_{c}}{(k+1)(l+1)},\\
\sigma_{DD,kl}&= \sigma^2_{b}\mathbf{1}_{\{k\not=0,l\not=0\}},\\
\sigma_{CD,kl}&=\left( \sigma^2_{b}-\frac{\sigma_{bc}}{k+1}\right)\mathbf{1}_{\{l\not=0\}},
\end{align}
\end{subequations}
for $k,l=0,1, \ldots, d-1$. Here, $\mathbf{1}_{A}$ is the indicator of an event $A$ defined by
\begin{equation}
\mathbf{1}_{A}= \left\{
    \begin{array}{ll}
      1& \mbox{if the event $A$ is true},\\
      0& \mbox{if the event $A$ is false}. 
    \end{array}
    \right.
\end{equation}
Then, we have
\begin{subequations}
\begin{align}
\mu_{C,k}-\mu_{D,k}&=\mu_{b}\mathbf{1}_{\{k=0\}}-\frac{\mu_{c}}{k+1},\\
\sigma_{CC,kl}-\sigma_{CD,kl}&=\left(\sigma^2_{b}-\frac{\sigma_{bc}}{k+1}\right)\mathbf{1}_{\{l=0\}}+\frac{\sigma^2_{c}}{(k+1)(l+1)}-\frac{\sigma_{bc}}{l+1},
\\
\sigma_{DD,kl}-\sigma_{CD,kl}&=-\sigma^2_{b}\mathbf{1}_{\{k=0,l\not=0\}}+\frac{\sigma_{bc}}{k+1}\mathbf{1}_{\{l\not=0\}}.
\end{align}
\end{subequations}
Inserting these values in the expression (\ref{mainresult}) and substitying it in (\ref{s2-eq10}), the average abundance of $C$ can be approximated as
\begin{equation}\label{snowdrift-eq1}
\mathbb{E}^{\delta}[X]\approx\frac{1}{2}+\frac{\delta N}{e\theta}\left(M_b\mu_{b}+M_c\mu_{C}+M_{bb}\sigma^2_b+M_{bc}\sigma_{bc}+M_{cc}\sigma^2_c\right),
\end{equation}
where 
\begin{subequations}\label{snowdrift-eq2}
\begin{align}
M_{b}&=\psi_{d+1}^1,\\
M_{c}&=-\sum_{k=0}^{d-1}\frac{\binom{d-1}{k}}{k+1}\psi_{d+1}^{k+1}=-\frac{1}{d} \sum_{k=0}^{d-1}\binom{d}{k+1}\psi_{d+1}^{k+1}\nonumber\\
&=-\frac{1}{d} \left(\sum_{k'=0}^{d}\binom{d}{k'}\psi_{d+1}^{k'}-\psi_{d+1}^0\right)=-\frac{1}{d}\left(\psi_{1}^0-\psi_{d+1}^0\right),\\
M_{bb}&=-\sum_{l=1}^{d-1}\binom{d-1}{l}\psi_{2d+1}^{l+1}-\sum_{k=0}^{d-1}\binom{d-1}{k}\psi_{2d+1}^{k+2}\nonumber\\
&=\psi_{2d+1}^1-\psi_{d+2}^1-\psi_{d+2}^2=\psi_{2d+1}^1-\psi_{d+1}^1,\\
M_{cc}&=-\sum_{k=0}^{d-1}\sum_{l=1}^{d-1}\frac{\binom{d-1}{l}}{l+1}\frac{\binom{d-1}{k}}{k+1}\psi_{2d+1}^{k+l+2}=-\frac{1}{d^2}\sum_{k=0}^{d-1}\sum_{l=1}^{d-1}\binom{d}{l+1}\binom{d}{k+1}\psi_{2d+1}^{k+l+2}\nonumber\\
&=-\frac{1}{d^2}\left(
 \sum_{k'=0}^{d}\sum_{l'=0}^{d}\binom{d}{k'}\binom{d}{l'}\psi_{2d+1}^{k'+l'}-2\sum_{k'=0}^{d}\binom{d}{k'}\psi_{2d+1}^{k'}+\psi_{2d+1}^{0}
\right)\nonumber\\
&=-\frac{1}{d^2}\left(\psi_{1}^{0}-2\psi_{d+1}^{0}+\psi_{2d+1}^{0}\right),\\
M_{bc}&=\sum_{k=0}^{d-1}\frac{\binom{d-1}{k}}{k+1}\psi_{2d+1}^{k+2}
+\sum_{k,l=0}^{d-1}\frac{\binom{d-1}{l}\binom{d-1}{k}}{l+1} \psi_{2d+1}^{k+l+2}
+\sum_{k=0}^{d-1}\sum_{l=1}^{d-1}\frac{\binom{d-1}{l}\binom{d-1}{k}}{k+1} \psi_{2d+1}^{k+l+1}\nonumber
\\
&=\frac{1}{d}\left[\sum_{k=0}^{d-1}\binom{d}{k+1}\psi_{2d+1}^{k+2}
+\sum_{k,l=0}^{d-1}\binom{d}{l+1}\binom{d-1}{k}\psi_{2d+1}^{k+l+2}
+\sum_{k=0}^{d-1}\sum_{l=1}^{d-1}\binom{d-1}{l}\binom{d}{k+1} \psi_{2d+1}^{k+l+1}\right]\nonumber\\
&=\frac{1}{d}\left[\sum_{k'=1}^{d}\binom{d}{k'}\psi_{2d+1}^{k'+1}
+\sum_{l'=1}^{d}\binom{d}{l'}\psi_{d+2}^{l'+1}
+\sum_{k'=1}^{d}\binom{d}{k'} \left(\psi_{d+2}^{k'}-\psi_{2d+1}^{k'}\right)\right]\nonumber\\
&=\frac{1}{d}\left[\psi_{d+1}^{1}-\psi_{2d+1}^{1}
+\psi_{2}^{1}-\psi_{d+2}^{1}
+\psi_{2}^{0}-\psi_{d+1}^{0}-(\psi_{d+2}^{0}-\psi_{2d+1}^{0})\right]\nonumber\\
&=\frac{1}{d}\left(\psi_{1}^{0}-2\psi_{d+1}^{0}+\psi_{d+1}^{1}+\psi_{2d+1}^{0}-\psi_{2d+1}^{1}\right).
\end{align}
\end{subequations}
In the calculation of these different coefficients, we have used $\binom{d-1}{k}/(k+1)=\binom{d}{k+1}/d$ and the identities (\ref{identity1}) and (\ref{identity2}).
Note that $M_{bb}<0$, $M_{cc}<0$ and $M_{bc}>0$, from which we can state the following result.
\paragraph{Result 5}\textit{
For the stag hunt game, decreasing the variance of the cost $c$, $\sigma^{2}_c$, or the variance of the benefit $b$, $\sigma^{2}_b$, or increasing their covariance, $\sigma_{bc}$, increases the average abundance of $C$ for any scaled mutation rate $\theta>0$ and any group size $d\geq2$.
}

Note that the condition for weak selection to favor the abundance of $C$ can be written as
\begin{align}
&\psi_{d+1}^1\mu_{b}+\frac{1}{d}\left(\psi_{1}^{0}-2\psi_{d+1}^{0}+\psi_{d+1}^{1}+\psi_{2d+1}^{0}-\psi_{2d+1}^{1}\right)\sigma_{bc}\nonumber\\
&-\frac{1}{d}\left(\psi_{1}^0-\psi_{d+1}^0\right)\mu_{C}
-\left(\psi_{d+1}^1-\psi_{2d+1}^1\right)\sigma^2_b
-\frac{1}{d^2}\left(\psi_{1}^{0}-2\psi_{d+1}^{0}+\psi_{2d+1}^{0}\right)\sigma^2_c>0.
\end{align}

\subsection{Large group size}
Another interesting point in the snowdrift game is the effect of the group size on the average abundance of $C$. 
Assuming a large population and using (\ref{stag-eq9}) and (\ref{stag-eq9bis}), we deduce that $M_b, M_{bb}, M_{cc} \approx o\left(d^{-1}\right)$, while
\begin{subequations}
\begin{align}
M_c&\approx-\frac{1}{2d}+o\left(d^{-1}\right),\\
M_{bc}&\approx \frac{1}{2d}+o\left(d^{-1}\right).
\end{align}
\end{subequations}
Inserting these expressions in (\ref{snowdrift-eq1}), the average abundance of $C$ can be approximated as
\begin{equation}
\mathbb{E}^{\delta}[X]\approx\frac{1}{2}+\frac{\delta N}{4d\theta}\left(\sigma_{bc}-\mu_{c}\right).
\end{equation}
We conclude that weak selection favors the abundance of $C$ as long as $\sigma_{bc}>\mu_c$. Note that, in this case, increasing the scaled mutation rate decreases the average abundance of $C$. We summarize these findings.

\paragraph{Result 6}\textit{
In the case of interactions in large groups in a large population, weak selection favors the abundance of $C$ as long as $\sigma_{bc}>\mu_c$. This is true for any scaled mutation rate $\theta>0$.}

\subsection{Low scaled mutation rate}
In this section and the next one, we study the effect of the scaled mutation rate in a large population on the average abundance of $C$ for any group size $d\geq2$.
Suppose first a low scaled mutation rate $\theta$.
Using (\ref{approximationsmall}), the different coefficients in (\ref{snowdrift-eq2}) can be approximated as
\begin{subequations}
\begin{align}
M_{b}&\approx \frac{\theta}{2d}+o(\theta),\\
M_{c}&\approx -\frac{\theta}{2d}H_d+o(\theta),\\
M_{bb}&\approx -\frac{\theta}{4d}+o(\theta),\\
M_{cc}&\approx -\frac{\theta}{2d^2}\left[2H_d-H_{2d}\right]+o(\theta),\\
M_{bc}&\approx \frac{\theta}{2d}\left[2H_d-H_{2d}+\frac{1}{2d}\right]+o(\theta).
\end{align}
\end{subequations}
Inserting these approximations in (\ref{snowdrift-eq1}), the average abundance of $C$ can be expressed as 
\begin{equation}
\mathbb{E}^{\delta}[X]\approx\frac{1}{2}+\frac{\delta N}{4d}\left[\mu_{b}-H_d\mu_{c}-\frac{\sigma^2_{c}}{2}-\frac{2H_d-H_{2d}}{d}\sigma^2_{b}+\left(2H_d-H_{2d}+\frac{1}{2d}\right)\sigma_{bc}\right].
\end{equation}
This means that weak selection favors the abundance of $C$ as long as
\begin{equation}
\mu_{b}-H_d\mu_{c}-\frac{\sigma^2_{c}}{2}-\frac{2H_d-H_{2d}}{d}\sigma^2_{b}+\left(2H_d-H_{2d}+\frac{1}{2d}\right)\sigma_{bc}>0.
\end{equation}
In the case of a large group size $d$, the above condition reduces
\begin{equation}
\sigma_{bc}>\mu_c
\end{equation}
owing to the approximation $H_d\approx\ln d$.

\subsection{High scaled mutation rate}
In the case of a high scaled mutation rate $\theta$ so that we have the approximation (\ref{approximationlarge}), the different coefficients in (\ref{snowdrift-eq1}) can be expressed as
\begin{subequations}
\begin{align}
M_{b}&\approx\frac{1}{2^{d+1}},\\
M_{c}& \approx-\frac{1}{d}\left(\frac{1}{2}-\frac{1}{2^{d+1}}\right),\\
M_{bb}&\approx\frac{1}{2^{2d+1}}-\frac{1}{2^{d+1}},\\
M_{cc}&\approx-\frac{1}{d^2}\left[\frac{1}{2}-\frac{1}{2^d}+\frac{1}{2^{2d+1}}\right],\\
M_{bc}&\approx\frac{1}{d}\left(\frac{1}{2}-\frac{1}{2^{d+1}}\right).
\end{align}
\end{subequations}
Then, the condition for weak selection to favor the abundance of $C$ becomes
\begin{align}
&\frac{1}{2^{d+1}}\mu_b-\frac{1}{d}\left(\frac{1}{2}-\frac{1}{2^{d+1}}\right)\mu_c-\left(\frac{1}{2^{d+1}}-\frac{1}{2^{2d+1}}\right)\sigma^2_{b}\nonumber\\
&-\frac{1}{d^2}\left[\frac{1}{2}-\frac{1}{2^d}+\frac{1}{2^{2d+1}}\right]\sigma^2_c+\frac{1}{d}\left(\frac{1}{2}-\frac{1}{2^{d+1}}\right)\sigma_{bc}>0,
\end{align}
which reduces again to
\begin{equation}
\sigma_{bc}>\mu_c
\end{equation}
in the case of a large group size $d$.

\section{Discussion}
In this paper, we have studied the average abundance of $C$ in the mutation-selection equilibrium of a finite well-mixed population, where interactions between cooperators represented by $C$ and defectors represented by $D$ occur in groups of size $d\geq2$. We have supposed random payoffs from one time step to the next in a discrete-time Moran model to reflect stochastic fluctuations in the environment. In the presence of recurrent mutation, we have shown that the average abundance of $C$ depends not only on the means of the payoffs but also on their second moments. This generalizes a previous study (Kroumi and Lessard \cite{KL2020}) in the case of pairwise interactions.

In addition to the expected scaled means, variances and covariances of the payoffs, the condition for weak selection to favor the abundance of $C$ is obtained in terms of identity measures. These are probabilities for a random sample of $n$ individuals in a neutral population at equilibrium to contain exactly $k$ cooperators, $\psi^{k}_n$, for $0\leq k\leq n$. In a large population, they correspond to moments of the Dirichlet distribution with scaled mutation rate $\theta \approx Nu/2$ where $N$ is the population size and $u$ the probability of mutation from one time step to the next. Their expressions have been deduced  in appendix $B$ using a coalescent approach developed in Griffiths and Lessard \cite{GL2005}.

When the payoffs are constant, we have shown that weak selection favors the abundance of $C$ if
\begin{equation}
\sum_{k=0}^{d-1}\binom{d-1}{k}\psi_{d+1}^{k+1}\mu_{C,k}>\sum_{k=0}^{d-1}\binom{d-1}{k}\psi_{d+1}^{k+1}\mu_{D,k}.
\end{equation}
Here, $\mu_{C,k}$ and $\mu_{D,k}$ are the scaled means of the payoffs to $C$ and $D$ in interaction with $k$ cooperators and $d-k-1$ defectors, respectively. This analytical result generalizes the condition obtained by Gokhale and Traulsen \cite{GT2011} for $d=3$. Our result is valid for any group size $d\geq 2$ and any scaled mutation rate $\theta > 0$ in the limit of a large population size. Note that the scaled mutation rate does not come into play in this condition for $d=2$. For $d\geq3$, the scaled mutation rate can enhance or lessen the condition for a strategy to be more abundant on average than another. In the limit of a low scaled mutation rate, weak selection favors the abundance of $C$ if 
\begin{equation}
\sum_{k=0}^{d-1}\mu_{C,k}>\sum_{k=0}^{d-1}\mu_{D,k}.
\end{equation}
 This is exactly the condition obtained by Kurokawa and Ihara \cite{KI2009} for weak selection to favor the evolution of $C$ more than the evolution of $D$ in the absence of mutation in a large population. 

In the case of random fluctuations in the payoffs, we have shown that the average abundance of $C$ exceeds the average abundance of $D$ if 
\begin{align}
&\sum_{k=0}^{d-1}\binom{d-1}{k}\psi_{d+1}^{k+1}\mu_{C,k}-\sum_{k,l=0}^{d-1}\binom{d-1}{k}\binom{d-1}{l}
\psi_{2d+1}^{k+l+2}\left(\sigma_{CC,kl}-\sigma_{CD,kl}\right)>\nonumber
\\
&\sum_{k=0}^{d-1}\binom{d-1}{k}\psi_{d+1}^{k+1}\mu_{D,k}-\sum_{k,l=0}^{d-1}\binom{d-1}{k}\binom{d-1}{l}
\psi_{2d+1}^{k+l+1}\left(\sigma_{DD,kl}-\sigma_{CD,kl}\right),
\end{align}
where $\sigma_{XY,kl}$ is the scaled covariance of the payoff to $X$ in interaction with $k$ cooperators and $d-k-1$ defectors and the payoff to $Y$ in interaction with $l$ cooperators and $d-l-1$ defectors, for $X,Y=C,D$. Moreover, we have shown that a decrease in the scaled covariance between any two payoffs to $C$, or an increase in the scaled covariance between any two payoffs to $D$, will increase the average abundance of $C$. Note that, at least in a large population, an increase in $\sigma_{CD,kl}$ will increase the average abundance of $C$ if $k+l>d-1$, and decrease the average abundance of $C$ if $k+l<d-1$.  Moreover, if $k+l=d-1$, then $\sigma_{CD,kl}$ does not have any effect on the average abundance of $C$.

These results are in agreement with previous studies on the effect of variability on the evolution of a trait based on fixation probability (Gillespie \cite{G1974}, Rychtar and Taylor \cite{RT2020}), Li and Lessard \cite{LL2020}), stochastic evolutionary stability (SES) and stochastic convergence stability (SCS) (Zheng \textit{et al.} \cite{ZLLT2017,ZLLT2018}), as well as stability concepts in a stochastic replicator equation (Imhof \cite{I2005}).

We have applied our results to different scenarios in the case of additive scaled mean cost and benefit for cooperation. With constant payoffs, weak selection favors the abundance of $D$ for any scaled mutation rate and any group size $d\geq2$. We have shown that introducing uncertainty in the payoffs to $D$ will increase the average abundance of $C$ and make it possible for weak selection to favor the abundance of $C$. This may be the case even if uncertainty is introduced in only one of the payoffs to $D$, e.g., the payoff $b_{k_0}$ when $D$ is in interaction with $k_0$ cooperators and $d-k_0-1$ defectors  (Case $1$ in Section $4$). In this case, we have shown that the abundance of $C$ is favored by weak selection if $\sigma^2/\mu_c>(\sigma^2/\mu_c)^*$, where  $\sigma^2$ is the scaled variance of $b_{k_0}$ while  all the other variances and covariances are insignificant. Nevertheless, the threshold $(\sigma^2/\mu_c)^*$ is increasing to infinity as $d$ increases to infinity for a scaled mutation rate that is low enough or high enough , so that it is not possible for weak selection to favor the abundance of $C$ if interactions occur in large enough groups. In the second scenario, where we suppose that the covariances of any two payoffs to $D$ are of the same magnitude given by $\sigma^2$ (Case $2$ in Section $4$),  we have shown that weak selection favors the abundance of $C$ in a large population if $\sigma^2/\mu_c>2$. This condition  does not depend neither on the scaled mutation rate nor on the group size. This means that it is possible for weak selection to favor the abundance of $C$ even if interactions occur in large groups which makes a difference from Case $1$.

Next point of interest is the abundance of $C$ in classical social dilemmas with random cost $c$ and benefit $b>c$ for cooperation in multi-player games. In the case of the public goods game, we have shown that the abundance of $C$ is favored by weak selection if 
\begin{equation}
\sigma_{bc}-\sigma^2_{c}>2\mu_c.
\end{equation}
Here, $\mu_c$ is the scaled expected cost, $\sigma^2_c$ the scaled variance of the cost, and $\sigma_{bc}$ the scaled covariance between the benefit $b$ and the cost $c$. In addition, we have shown that a decrease in $\sigma^2_c$, or an increase in $\sigma_{bc}$, will increase the average abundance of $C$.
The above condition does not depend neither on the scaled mutation rate nor on the group size. Note, however, that an increase in the scaled mutation rate can increase or decrease the average abundance of $C$.

The stag hunt game is when cooperation implies a cost $c$ to receive a benefit $b$ if all individuals cooperate.
 In such a case, We have shown that a decrease in the scaled variance of the cost $c$ or the benefit $b$, or an increase in their scaled covariance, will increase the average abundance of $C$. Increasing the group size  $d$ will reduce the weight of $\sigma_{bc}$, which makes it more difficult for weak selection to favor the abundance of $C$. With interactions in large groups, weak selection will never favor the abundance of $C$, which is true for any mutation probability. This is a consequence of the fact that  a cooperator will receive the benefit if all its partners cooperate, which will occur rarely if $d$ is large. 
 
The snowdrift game is when the cost is distributed equally between all cooperators in the group. In such a scenario, all the conclusions obtained in the stag hunt game are still valid except the effect of increasing the group size. In the snowdrift game and in the case of interactions in large groups, weak selection favors the abundance of $C$ as long as $\sigma_{bc}>\mu_c$ for any mutation probability. This condition does not depend on $\sigma^2_{c}$, contrary to the case in the public goods game.

\section*{Funding}
D. Kroumi is funded by Deanship of Scientific Research (DSR) at King Fahd University of Petroleum and Minerals (GRANT SR181014). S. Lessard is supported by NSERC of Canada, grant  no. 8833.

\section*{Acknowledgments}
D. Kroumi would like to acknowledge the support provided by the Deanship of Scientific Research (DSR) at King Fahd University of Petroleum and Minerals (KFUPM) for funding this work. S. Lessard is supported 
by NSERC of Canada, grant no. 8833.


\section{Appendix A: Conditional expected frequency change}
For $i/N=x$, we have
\begin{equation}
\begin{split}
\frac{\binom{i-1}{k}\binom{N-i}{n-k}}{\binom{N-1}{n}}&=\binom{n}{k}\frac{(i-1)\cdots(i-k)\times(N-i)\cdots(N-i-n+k+1)}{(N-1)\cdots(N-n)}\\
&=\binom{n}{k}\frac{(x-\frac{1}{N})\cdots(x-\frac{k}{N})\times(1-x)(1-x-\frac{1}{N})\cdots(1-x-\frac{n-k-1}{N})}{(1-\frac{1}{N})\cdots(1-\frac{n}{N})}\\
&=\binom{n}{k}x^{k}(1-x)^{n-k}+O(N^{-1}).
\end{split}
\end{equation}
Therefore, in the limit of a large population size $N$, the average payoffs to $C$ and $D$ in (\ref{payoffsCD}) can be written as
\begin{subequations}
\begin{align}
P_C(x)&=\sum_{k=0}^{d-1}\binom{d-1}{k}x^k(1-x)^{d-k-1}a_k,\\
P_D(x)&=\sum_{k=0}^{d-1}\binom{d-1}{k}x^k(1-x)^{d-k-1}b_k.
\end{align}
\end{subequations}
Using ( \ref{momentspayoffs}), the first two moments of $P_C(x)$ are given by
\begin{subequations}
\begin{align}
E\Big[P_C(x)\Big]&=\delta\sum_{k=0}^{d-1}\binom{d-1}{k}x^k(1-x)^{d-k-1}\mu_{C,k}+o(\delta)\label{A-eq1},\\
E\Big[P_C^2(x)\Big]&=\delta\sum_{k,l=0}^{d-1}\binom{d-1}{k}\binom{d-1}{l}x^{k+l}(1-x)^{2d-k-l-2}\sigma_{CC,kl}+o(\delta)\label{A-eq2},
\end{align}
\end{subequations}
and the first two moments of $P_D(x)$ by
\begin{subequations}\label{A-eq3}
\begin{align}
E\Big[P_D(x)\Big]&=\delta\sum_{k=0}^{d-1}\binom{d-1}{k}x^k(1-x)^{d-k-1}\mu_{D,k}+o(\delta),\\
E\Big[P_D^2(x)\Big]&=\delta\sum_{k,l=0}^{d-1}\binom{d-1}{k}\binom{d-1}{l}x^{k+l}(1-x)^{2d-k-l-2}\sigma_{DD,kl}+o(\delta).
\end{align}
\end{subequations}
Moreover, we have
\begin{align}\label{A-eq4}
E\Big[P_C(x)P_D(x)\Big]=\delta\sum_{k,l=0}^{d-1}\binom{d-1}{k}\binom{d-1}{l}x^{k+l}(1-x)^{2d-k-l-2}\sigma_{CD,kl}+o(\delta).
\end{align}
We are interested in
\begin{equation}
E\left[\frac{f_C(x)-f_D(x)}{xf_C(x)+(1-x)f_D(x)}\right]=E\left[\frac{P_C(x)-P_D(x)}{1+\bar{P}(x)}\right],
\end{equation}
where $\bar{P}(x)=xP_C(x)+(1-x)P_D(x)$ is the average payoff in the population.
We expand the last expression by the delta-method (Lynch and Walsh \cite{LW1998}, Rice and Papadopoulos \cite{RP2009}), which gives 
\begin{equation}
E\left[\frac{Y}{Z}\right]=\frac{E[Y]}{E[Z]}+\sum_{k=1}^{\infty}(-1)^k\frac{E[Y]\ll \prescript{k}{}{Z}\gg+\ll Y,\prescript{k}{}{Z}\gg}{E[Z]^{k+1}}
\end{equation}
for two random variables $Y$ and $Z$ with
\begin{align}
\ll \prescript{k}{}{Z}\gg =E\left[\left(Z-E[Z]\right)^k\right]
\end{align} 
and
\begin{align}
\ll Y,\prescript{k}{}{Z}\gg =E\left[\left(Y-E[Y]\right)\left(Z-E[Z]\right)^k\right]
\end{align} 
for $k \geq 1$.
For $Y=P_C(x)-P_D(x)$ and $Z=1+\bar{P}(x)$, using the facts that 
\begin{align}
\ll P_C(x)-P_D(x),\prescript{k}{}{(1+\bar{P}(x))}\gg=\ll P_C(x)-P_D(x),\prescript{k}{}{\bar{P}(x)}\gg=o(\delta)
\end{align}
for $k\geq 2$ and
\begin{align}
E[P_C(x)-P_D(x)]\ll \prescript{k}{}{(1+\bar{P}(x))}\gg=E[P_C(x)-P_D(x)]\ll \prescript{k}{}{\bar{P}(x)}\gg=o(\delta),
\end{align}
for $k\geq1$,
we obtain
\begin{equation}
E\left[\frac{P_C(x)-P_D(x)}{1+\bar{P}(x)}\right]=\frac{E[P_C(x)-P_D(x)]}{1+E[\bar{P}(x)]}-\frac{\ll P_C(x)-P_D(x),\prescript{1}{}{\bar{P}(x)}\gg}{(1+E[\bar{P}(x)])^{2}}+o(\delta).
\end{equation}
Note that 
\begin{align}
\ll P_C(x)-P_D(x),\prescript{1}{}{\bar{P}(x)}\gg&=E\left[\left(P_C(x)-P_D(x)\right)\bar{P}(x)\right]+E\left[P_C(x)-P_D(x)\right]E\left[\bar{P}(x)\right]\nonumber\\
&=E\left[\left(P_C(x)-P_D(x)\right)\bar{P}(x)\right]+o(\delta),
\end{align}
and $\left(1+E[\bar{P}(x)]\right)^{2}=1+O(\delta)$, so that
\begin{align}
E\left[\frac{P_C(x)-P_D(x)}{1+\bar{P}(x)}\right]&=E[P_C(x)]-E[P_D(x)] -E\left[\left(P_C(x)-P_D(x)\right)\bar{P}(x)\right]+o(\delta)\nonumber\\
&=E[P_C(x)]-E[P_D(x)]-xE[P^2_1(x)]+(2x-1)E[P_C(x)P_D(x)]\nonumber\\
&\quad+(1-x)E[P^2_2(x)]+o(\delta)\nonumber\\
&=m(x)\delta+o(\delta),
\end{align}
where 
\begin{align}
m(x)&=\sum_{k=0}^{d-1}\binom{d-1}{k}x^k(1-x)^{d-k-1}\left(\mu_{C,k}-\mu_{D,k}\right)+\sum_{k,l=0}^{d-1}\binom{d-1}{k}\binom{d-1}{l}\nonumber\\
&\times x^{k+l}(1-x)^{2d-k-l-2}\Big[-x\sigma_{CC,kl}+(1-x)\sigma_{DD,kl}+((x-(1-x))\sigma_{CD,kl}\Big].
\end{align}


\section{Appendix B: Moments of the Dirichlet distribution}
 
The ancestry of a random sample of $n$ individuals is described backward in time by a coalescent tree  with every pair of lines at any given time back coalescing at rate $1$ independently of all the others (Kingman \cite{K1982}). Moreover, mutations occur independently on the lines of the coalescent tree according to a Poisson process of intensity $\theta >0$. When there is mutation, the mutant type is $1$ or $2$ with probability $1/2$ for each type independently of the parental type. A line is said to be ancestral to the sample as long as no mutation has occurred on it.
We are interested in the probability distribution of the sample configuration, more precisely the probability for $k$ labeled individuals to be of type $1$ and the $n-k$ others to be of type $2$, denoted by $\psi_n^k$, for $0\leq k\leq n$. 

In order to determine the sample probability distribution, we will extend an approach used in  Griffiths and Lessard \cite{GL2005} to show the Ewens sampling formula  (Ewens \cite{E1972}) in the case where the mutant type is always a novel type. See also Hoppe \cite{H1984} and  Joyce and Tavar\'e \cite{JT1987} for related approaches based on urn models and cycles in permutations.

Note first that the number of ancestral lines of a sample of $n$ individuals is a death
process backward in time, where ancestral lines are lost by either mutation or
coalescence. This death process was studied in Griffiths \cite{G1980} and Tavar\'e \cite{T1984}, and the events in this death process called defining events in Ewens \cite{E1990}.

Label the $n$ sampled individuals and list them in the order in which their ancestral lines are
lost backward in time, following either a mutation or a
coalescence. In the case of coalescence, one of the two lines
involved is chosen at random to be the one
that is lost, the other one being the continuing line. There are $n!$ different orders.

Let us first consider the probability for the $n$ sampled individuals in a given order to be all of type $1$.
Note that this event occurs if and only if all ancestral lines lost by mutation lead to type $1$.
Now let us look at the probability of each defining
event. When $i$ ancestral lines remain, the total rate of mutation
is $i\theta$ and the total rate of coalescence is $i(i-1)/2$, while the rate of mutation leading to type $1$ on any particular ancestral line is $\theta/2$  and the rate of coalescence involving any particular ancestral line and leading to its loss is $(i-1)/2$.
Therefore, the probability that a particular ancestral line is the next one lost
and that it is lost by mutation leading to type $1$ is $(\theta/2)/[(i\theta+i(i-1)/2]$ for $i\geq 1$.
Similarly, the probability that a particular ancestral line is the next one lost and that it is lost by coalescence is
$[(i-1)/2]/[(i\theta+i(i-1)/2]$ for $i\geq1$. Summing these probabilities, multiplying the sums for $i=n, n-1, \ldots, 1$, and considering all possible orders, we get
\begin{align}
\psi_n^0&=n! \, \prod_{i=1}^{n}
\left(\frac{\theta/2 + (i-1)/2}{i\theta +i(i-1)/2} \right)
= \frac{\prod_{i=1}^{n}(\theta + i-1)}{ \prod_{i=1}^{n}(2\theta +i-1)}
 \end{align}
as probability for the $n$ sampled individuals to be of type $1$.

Now let us look at the general case of $k$ labeled individuals of type $1$ and $n-k$ of type $2$. When $i$ ancestral lines of individuals of type $1$ and $j$ ancestral lines of individuals of type $2$ remain, the total rate of mutation
is $(i+j)\theta$ and the total rate of coalescence is $(i+j)(i+j-1)/2$, while any particular ancestral line of an individual is lost by mutation to the type of the individual, whose rate is $\theta/2$, or by coalescence with ancestral lines  of individuals of the same type, whose rate is $(i-1)/2$ for type $1$ and $(j-1)/2$ for type $2$. Considering $i$ and $j$ decreasing from $k$ and $n-k$ to $0$ or $1$ with $i+j=n, n-1, \ldots, 1$, we get
\begin{align}
\psi_n^k&=n! \, \frac{\prod_{i=1}^{k}(\theta/2 + (i-1)/2) \prod_{j=1}^{n-k}(\theta/2 + (j-1)/2)}{ \prod_{l=1}^{n}(l\theta +l(l-1)/2)} \nonumber\\
&=\frac{\prod_{i=1}^{k}(\theta+i-1)\prod_{j=1}^{n-k}(\theta+j-1)}{\prod_{l=1}^{n}(2\theta+l-1)}\nonumber\\
&=\frac{\Gamma(2\theta)}{ \Gamma(2\theta +n) } \left(\frac{\Gamma(\theta +k)\Gamma(\theta +n-k)}{ \Gamma(\theta)^2}\right)
\end{align}
as probability for $k$ labeled individuals to be of type $1$ and the $n-k$ others to be of type $2$ in a random sample of size $n$. Here, we use $\Gamma(\beta +1  )= \beta \Gamma(\beta)$ for $\beta >0$. Note that $\psi_n^k=\psi_{n-k}^k$.

Applying the same approach as above for $K$ types with rate of mutation to type $k$ given by $\alpha_k/2 >0$ for $k=1, \ldots, K$ with $\alpha_1 + \cdots + \alpha_K=\alpha$, the probability for $n_k$ labeled individuals to be of type $k$ for $k=1, \ldots, K$ in a random sample of size $n=n_1 + \cdots + n_K$ is given by
\begin{align}
\psi_n^{n_1, \ldots, n_K}&= \frac{\prod_{k=1}^{K}\prod_{i_k=1}^{n_k}(\alpha_k + i_k-1)}{ \prod_{i=1}^{n}(\alpha +i -1)} \nonumber\\
&=\frac{\Gamma(\alpha)}{\Gamma(\alpha + n)}\prod_{k=1}^{K} \frac{\Gamma(\alpha_k + n_k)}{ \Gamma(\alpha_k)}.
\end{align}
These are the moments of the Dirichlet distribution (see, e.g., Balakrishnan and Nevzorov \cite{BN2003}).

\bibliographystyle{unsrt}

\end{document}